\documentclass{PoS}

\usepackage{amsmath}
\usepackage{amsfonts,amssymb}
\usepackage{bm}% bold math

\usepackage{dsfont}
\usepackage{slashed}
\usepackage{booktabs}

\newcommand{\be}{\begin{equation}}
\newcommand{\ee}{\end{equation}}
\newcommand{\bea}{\begin{eqnarray}}
\newcommand{\eea}{\end{eqnarray}}

\newcommand{\llangle}{\langle\hspace*{-0.08cm}\langle}
\newcommand{\rrangle}{\rangle\hspace*{-0.08cm}\rangle}

\newcommand{\Op}{\mathcal{O}} % Fractur O 
\newcommand{\PO}{\mathcal{P}} % path ordering operator                                                
\newcommand{\eins}{\mathds{1}} % Fractur O                                           

\newcommand{\Dlr}{\buildrel \leftrightarrow \over D\raise-1pt\hbox{}}

\title{Nucleon transversity generalized form factors with twisted mass fermions}

\ShortTitle{Nucleon transversity generalized form factors}

\author{C. Alexandrou\\
         Department of Physics, University of Cyprus, P.O. Box 20537, 1678 Nicosia, Cyprus, and\\  
 Computation-based Science and Technology Research  
    Center, Cyprus Institute, 20 Kavafi Str., Nicosia 2121, Cyprus \\  
        E-mail: \email{alexand@ucy.ac.cy}}
\author{\speaker{M. Constantinou}\\
         Department of Physics, University of Cyprus, P.O. Box 20537, 1678 Nicosia, Cyprus\\
       E-mail: \email{marthac@ucy.ac.cy}}
\author{K. Jansen\\
         NIC, DESY, Platanenallee 6, D-15738 Zeuthen, Germany\\
        E-mail: \email{karl.jansen@desy.de}}
\author{G. Koutsou\\
Computation-based Science and Technology Research  
 Center, Cyprus Institute, 20 Kavafi Str., Nicosia 2121, Cyprus\\
        E-mail: \email{g.koutsou@cyi.ac.cy}}
\author{H. Panagopoulos\\
          Department of Physics, University of Cyprus, P.O. Box 20537, 1678 Nicosia, Cyprus\\
        E-mail: \email{haris@ucy.ac.cy}}

\abstract{We present results on the nucleon tensor form factors and
  first moment of the transversity distribution using maximally
  twisted mass fermions. We analyze two $N_f{=}2{+}1{+}1$ ensembles
  having pion masses of 213~MeV and 373~MeV with lattice spacing
  $a=0.064$~fm and $a=0.082$~fm, respectively. First results using an
  $N_f{=}2$ ensemble of twisted mass fermions with a clover term at a
  physical pion mass and lattice spacing $a=0.094$~fm are also
  presented. The renormalization function for the local tensor form
  factors is evaluated non-perturbatively with a perturbative
  subtraction of ${\cal O}(a^2)$-terms, while for the first moment of
  the transversity we use a perturbative estimate. Results are given
  in the $\overline{\rm MS}$ scheme at a scale of $ \mu=2$~GeV, and
  are compared with recent results obtained using different
  discretization schemes.}

\FullConference{31st International Symposium on Lattice Field Theory - LATTICE 2013\\
		July 29 - August 3, 2013\\
		Mainz, Germany}

\begin{document}

\section{Introduction}

Lattice QCD calculations of observables related to the structure of
baryons are now being carried out using simulations of the theory with
pion mass close or even at the physical value~\cite{Hagler:2007xi, Syritsyn:2009mx, Brommel:2007sb,Yamazaki:2009zq,Alexandrou:2013joa}.
Nucleon observables that are under intense experimental study are 
the Generalized Parton Distributions (GPDs), which encode important
information on nucleon structure.  The GPDs can be accessed
in high energy processes where QCD factorization applies, and 
 the amplitude can be written in terms of the convolution 
of a hard perturbative kernel
with the GPDs. The twist-2 GPDs, which are
studied in this paper, are defined by the matrix element:

\be
  F_{\Gamma}(x,\xi,q^2) =   \frac{1}{2}\int \! \frac{d\lambda}{2\pi}
     e^{ix\lambda} \langle p^\prime |\bar {\psi}(-\lambda n/2) {\Gamma}
\PO      e^{ig\!\int \limits_{-\lambda /2}^{\lambda /2}\! d\alpha\, n \cdot A(n\,\alpha)}
     \psi(\lambda n/2) |p\rangle \,,\label{bilocal}
\ee
\vspace*{-0.2cm}

\noindent
where $|p^\prime\rangle $ and $|p\rangle$ are one-particle states,
$q=p^\prime-p$, $\xi=-n\cdot q/2$, $x$ is the momentum fraction, and
$n$ is a light-like vector collinear to $P=(p+p')/2$ and such that
${P}\cdot n=1$. The gauge link $\PO \exp(\dots) $ is necessary for
gauge invariance. In model calculations it is often set to one, which
amounts to working with QCD in the light-like gauge $A\cdot n=0$, but
on the lattice such a gauge fixing is not necessary. In the forward
limit, for which $\xi=0$ and $q^2=0$, GPDs reduce
to the ordinary parton distributions, namely the longitudinal momentum,
$q(x)$, the helicity, $\Delta q(x)$, and transversity, $\delta q(x)$,
distributions; in this paper we restrict to the transverity which
represents the net number of quarks with transverse polarization in a
transversely polarized nucleon. The first few Mellin moments of the
transversity parton distribution are of particular interest
\begin{eqnarray}
\langle x^n\rangle_{\delta q} = \int_{0}^{1}dx \, x^n\left[\delta
  q(x)+(-1)^{n+1}\delta\bar{q}(x)\right] \>\,\,, \qquad
\delta q=q_\top+q_\bot \,.
\end{eqnarray}
\noindent
The matrix elements of the light-cone operator as defined
in Eq.~(\ref{bilocal}) cannot be extracted from correlators in
euclidean lattice QCD but an operator
product expansion can be carried out leading to
\be
\Op_\top^{\mu\nu\mu_1\ldots\mu_{n-1}}=  \bar q
\,{{\sigma}}^{[\mu\,\{\nu]}\,i D^{\mu_1}\ldots i
D^{\mu_{n-1}\}} q\,.
\label{Oper_def}
\ee
\noindent
The curly brackets represent a symmetrization over indices and
subtraction of traces, while the square brackets represent
antisymmetrization over indices. Here we study the cases $n=0,\,1$, which amount
to calculating the local and one-derivative tensor currents,
respectively. The matrix elements of these operators are
parameterized in terms of the generalized form factors (GFFs)
$A_{T10},\,B_{T10},\,\widetilde A_{T10}$ and
$A_{T20},\,B_{T20},\,\widetilde A_{T20},\,\widetilde B_{T21}$
depending only on $q^2=(p^\prime-p)^2$:
\bea
\llangle \overline q(0)i\sigma^{\mu\nu} q(0)\rrangle &=&
\llangle i\sigma^{\mu\nu}\rrangle\, A_{T10}(q^2) \,+
\llangle\frac{\gamma^{[\mu}\Delta^{\nu]}} {2 m_N}\rrangle
\,B_{T10}(q^2)
+\llangle\frac{\overline P^{[\mu} \Delta^{\nu]}} {m_N^2}\rrangle\,\widetilde A_{T10}(q^2)\,, \\
\llangle \overline q(0)
\mathcal{O}_T^{\mu\nu\mu_1}(0) q(0)\rrangle \hspace*{-0.1cm}&=&\hspace*{-0.1cm}
\mathcal{A}_{\mu\nu}\mathcal{S}_{\nu\mu_1}  \bigg\{
\llangle i\sigma^{\mu\nu} \overline P^{\mu_1}\rrangle\, A_{T20}(q^2)
+ \llangle\frac{\gamma^{[\mu}\Delta^{\nu]}} {2 m_N} \overline
P^{\mu_1}\rrangle \,B_{T20}(q^2)\nonumber\\
&+&   \llangle\frac{\overline P^{[\mu} \Delta^{\nu]}} {m_N^2 }
\overline P^{\mu_1}\rrangle\, \widetilde A_{T20}(q^2)
+ \llangle\frac{\gamma^{[\mu}\overline P^{\nu]}} {m_N}
\Delta^{\mu_1}\rrangle \, \widetilde B_{T21}(q^2) \bigg\}\,.
\eea
\vskip -0.2cm
\noindent
In the forward limit we can directly obtain
$A_{T10}(0)=\langle 1 \rangle_{\delta q(x)} $ and 
$A_{T20}(0)=\langle x\rangle_{\delta q(x)}$.

\section{Evaluation on the lattice}

In the present work we employ the twisted mass fermion (TMF)
action with $N_f{=}2{+}1{+}1$ dynamical quarks~\cite{Baron} and the Iwasaki improved
gauge action. We also present results for an ensemble of $N_f{=}2$
TMFs with a clover term and tree-level Symanzik
gauge action. Using 
standard techniques, the GFFs are extracted from dimensionless ratios of
correlation functions, involving two-point and three-point functions:
\bea
G(\vec q, t_f-t_i)&=&
\sum_{\vec x_f} \, e^{-i(\vec x_f-\vec x_i) \cdot \vec q}\,
     {\Gamma^0_{\beta\alpha}}\, \langle {J_{\alpha}(t_f,\vec
       x_f)}{\overline{J}_{\beta}(t_i,\vec{x}_i)} \rangle\,, \label{twop}\\
G^{\mu\nu\mu_1}(\Gamma^k,\vec q,t) &=&
\sum_{\vec x, \vec x_f}e^{i(\vec x -\vec x_i)\cdot \vec q}\,
\Gamma^k_{\beta\alpha}\, \langle
{J_{\alpha}(t_f,\vec x_f)} \Op^{\mu\nu\mu_1}(t,\vec x)
{\overline{J}_{\beta}(t_i,\vec{x}_i)}\rangle \>\,.\label{threep}
\eea
\vskip -0.2cm
\noindent
 We consider
kinematics for which the final momentum $\vec{p}^\prime=0$ and 
we employed the fixed-sink method which requires a fixed time
separation between the sink and the source, $t_f-t_i$. The projection
matrices ${\Gamma^0}$ and ${\Gamma^k}$ are given by
\vspace*{-0.3cm}
\be
{\Gamma^0} = \frac{1}{4}(\eins + \gamma_0)\,,\quad 
\sum_{k=1}^3 {\Gamma^k} ={\Gamma^0} i \gamma_5 \sum_{k=1}^3 \gamma_k \, .\label{proj}
\ee
\vskip -0.2cm
\noindent
We use the standard proton interpolating field with Gaussian smeared
quark fields to increase the overlap with the proton state and
decrease overlap with excited states. We also apply APE-smearing to
the gauge fields $U_\mu$~\cite{Alexandrou:2013joa}. For matrix elements of
 isovector 
operators the disconnected contributions are zero up to lattice artifacts. For
the isoscalar local tensor we have computed the disconnected diagram,
which was found to be very small~\cite{Abdel-Rehim:2013wlz}.
We form an appropriate ratio of three- and two- functions
\be
R^{\mu\nu}(\Gamma^k,\vec q,t)=
\frac{G^{\mu\nu}(\Gamma^k,\vec q,t) }{G(\vec 0,
  t_f-t_i)} \,\times\, \sqrt{\frac{G(\vec p, t_f{-}t)G(\vec 0,
    t-t_i)G(\vec0,
    t_f-t_i)}{G(\vec 0  , t_f{-}t)G(\vec p,t-t_i)G(\vec
    p,t_f-t_i)}}\>,
\label{ratio}
\ee
\vskip -0.2cm
\noindent
which is optimized because it does not contain potentially noisy
two-point functions at large separations and because correlations
between its different factors reduce the statistical noise.
For sufficiently large time separations of the source and the sink,
this ratio becomes time-independent:
\vspace*{-0.25cm}
\be
\lim_{t_f-t\rightarrow \infty}\lim_{t-t_i\rightarrow  \infty}R^{\mu\nu}(\Gamma^\lambda,\vec q,t)=\Pi^{\mu\nu}(\Gamma^\lambda,\vec q) \,.
\label{plateau}
\ee

\begin{figure}[h!]
\begin{minipage}{0.37\linewidth}\vspace*{-.25cm}
{\includegraphics[width=\linewidth,angle=-90]{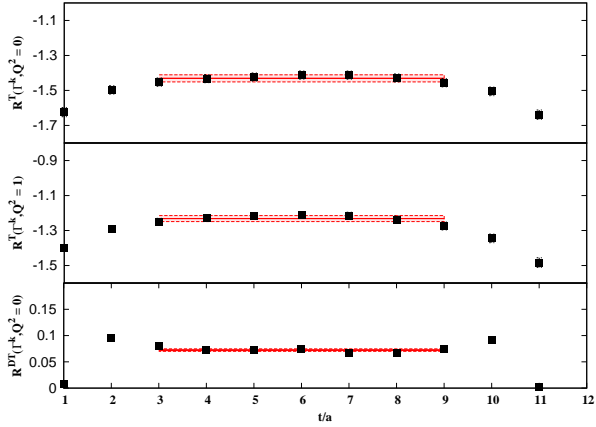}}
\caption{$R^{\mu\nu}$  (upper two) and $R^{\mu\nu\mu_1}$
 for representative choices of the momentum. The solid lines with
     the bands indicate the fitted ranges and plateau values with their jackknife
     errors.}\label{fig:plateaus}
\end{minipage}\hfill
\begin{minipage}{0.45\linewidth}\vspace*{-.5cm}
In this study we use sequential propagators already produced for the computation
of other nucleon matrix elements with $t_f-t_i\sim 1$~fm
namely, for the $N_f{=}2{+}1{+}1$ TMF ensembles we use
$(t_f-t_i)/a{=}12$ for $\beta=1.95$, $(t_f-t_i)/a{=}18$ for $\beta=2.10$
and for the $N_f{=}2$ TMF with a clover term ensemble, $(t_f-t_i)/a{=}12,\,14$.
For the latter ensemble we find that the 
results are compatible within error bars with the data for
$(t_f-t_i)/a{=}14$ carrying larger statistical errors. Thus, in the plots we
only show the results for $(t_f-t_i)/a{=}12$.
\end{minipage}
\end{figure}

From the plateau values of the renormalized asymptotic ratio
$\Pi(\Gamma^k, \vec q)_R=Z\,\Pi(\Gamma^kj, \vec{q})$
the nucleon matrix elements of the operators can be extracted.
All values of $\vec q$ corresponding to the same $q^2$, the two choices
of projector matrices $\Gamma^0$ and $\sum_k \Gamma^k$ and the
relevant orientations $\mu,\nu,\rho$ of the operators lead to an
over-constrained system of equations, which is solved in the
least-squares sense via a singular value decomposition of the
coefficient matrix. All quantities will be given in Euclidean space
with $Q^2\equiv-q^2$ being the Euclidean momentum transfer squared.
Both projectors $\Gamma^0$ and $\sum_k\Gamma^k$ are required to obtain
all GFFs at non-zero momentum. Not all combinations of the indices
$\mu,\,\nu,\,\mu_1$, are nessecary but we use all possibilities in
order to increase statistics. In Fig.~\ref{fig:plateaus} we show
representative plateau for the ratios of the local tensor and
the one derivative tensor operators at $\beta=1.95$, for different $\vec{Q}$-components.

\section{Renormalization}

\begin{figure}[h]
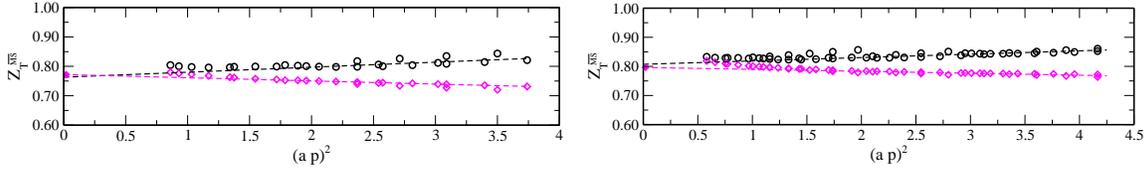

\begin{minipage}{0.49\linewidth}
\includegraphics[width=\linewidth]{Zt_MS_2GeV_b1.95.eps}
\end{minipage}\hfill
\begin{minipage}{0.49\linewidth}
\includegraphics[width=\linewidth]{Zt_MS_2GeV.eps}
\end{minipage}
\caption{$Z_T^{\overline{\rm MS}}(2\, {\rm GeV})$ for $N_f{=}4$ at $\beta=1.95$ (left) and
  $\beta=2.10$ (right). Black circles are the unsubtracted data
  and the magenta diamonds the data after subtracting the perturbative
  ${\cal O}(a^2)$-terms. The solid diamond at $(a\,p)^2=0$ is the
  value obtained after performing a linear extrapolation on the
  subtracted data.}
\label{Z_Nf4}
\end{figure}

We determine the necessary renormalization functions for the local
tensor operator non perturbatively in the RI$'$ scheme by
employing a momentum source at the vertex~\cite{Gockeler:1998ye},
which leads to high statistical accuracy and the evaluation of the
vertex for any operator at no significant additional computational
cost. For the details of the non-perturbative renormalization see
Ref.~\cite{Alexandrou:2010me}. In the RI$'$ scheme the renormalization
functions are determined in the chiral limit. For the renormalization of our
$N_f{=}2{+}1{+}1$ ensembles, ETMC has generated $N_f{=}4$ ensembles at the
same $\beta$ values, so that the chiral limit can be taken. To improve
our final estimates obtained from the continuum extrapolation we have
also computed the Green's functions related to the renormalization
functions in perturbation theory up to ${\cal O}(a^2)$
terms~\cite{Constantinou:2009tr,Alexandrou:2012mt};
we perform a perturbative subtraction of these ${\cal O}(a^2)$-terms.
This subtracts the leading cut-off effects yielding, in general, a
weak dependence of the renormalization functions on $(ap)^2$ and
the $(ap)^2\rightarrow 0$ limit can be reliably taken; this can be
seen in Fig.~\ref{Z_Nf4} for the two $N_f=2{+}1{+}1$ ensembles. As an
example, we present the perturbative terms that we subtract for the
Iwasaki gluonic action and clover coefficient $c_{\rm sw}=0$:

\be
a^2\,\frac{g^2\,C_F}{16\,\pi^2}\,\Bigg[ 0.2341\,\mu^2
+     \frac{8}{3}\,\frac{\mu 4}{\mu^2} + \log
(a^2\,\mu^2)\,\left(\frac{7271}{60000} \,\mu^2 - \frac{28891}{30000}\,
\frac{\mu4}{\mu^2} \right) \Bigg]\,,\quad \left(\mu 4\equiv \sum_{i=1,4}\mu_i^4 \right)\,. \nonumber
\ee
\vskip -0.1cm
\noindent
For the renormalization functions of the one-derivative tensor
operator, $Z_{\rm DT}$, we use our perturbative results~\cite{Alexandrou:2010me},
which we compute for general action parameters. For Iwasaki gluons
the expression for $Z_{\rm DT}$ in the RI$'$ scheme is:
\vspace*{-0.2cm}
\be
Z_{DT}(p=\bar\mu) =  1 + \frac{g^2\,C_F}{16\,\pi^2}\Biggl (2.3285 -
   2.2795\,c_{\rm sw}- 1.0117\,c_{\rm sw}^2 - 3\,\log \left(a^2\,\bar\mu^2\right)
   \Biggr)\,.
\ee
The renormalization functions are converted to the ${\overline{\rm MS}}$ scheme
at a scale of $\mu=2$~GeV using the conversion factors of Refs.~\cite{Gracey:2003yr,Gracey:2006zr}.
For the non-perturbative estimate of $Z_T$ we first subtract the
${\cal O}(a^2)$ perturbative terms and then apply the conversion to
the ${\overline{\rm MS}}$ scheme. The values of $Z_T^{\overline{\rm
    MS}}(2\, {\rm GeV})$ which we use in this paper are given below, where
the numbers in the parenthesis correspond to the statistical error.
As mentioned earlier, we use our perturbative results on $Z_{DT}$ to renormalize the traversity moment:
\vspace*{-0.2cm}
\bea
\beta=1.95,\,\, N_f=2+1+1 &:& Z_T = 0.625(2),\quad Z_{DT} = 1.019
\nonumber \\
\beta=2.10,\,\, N_f=2+1+1 &:& Z_T = 0.664(1),\quad Z_{DT} = 1.048
\nonumber \\
\beta=2.10,\,\, N_f=2, c_{sw}=1.58   &:& Z_T = 0.914(1),\quad Z_{DT} = 1.104
\nonumber
\eea
\vskip -0.1cm

\section{Lattice Results}

In this section we present results for the isovector and isoscalar
nucleon tensor charge $g_T \equiv A_{T10}(0)$, the first moment of
the transversity $<x>_{\delta q} \equiv A_{T20}(0)$, and compare with
results using other lattice discretizations. The renormalization
functions for the isoscalar quantities receive a contribution from a
disconnected diagram. For the Wilson gluonic action, the correction
was computed perturbatively and found to be very
small~\cite{Skouroupathis:2008mf}. We assume that the correction is
also small for the gauge action used here and it is therefore
neglected. 

\begin{figure}[h ]
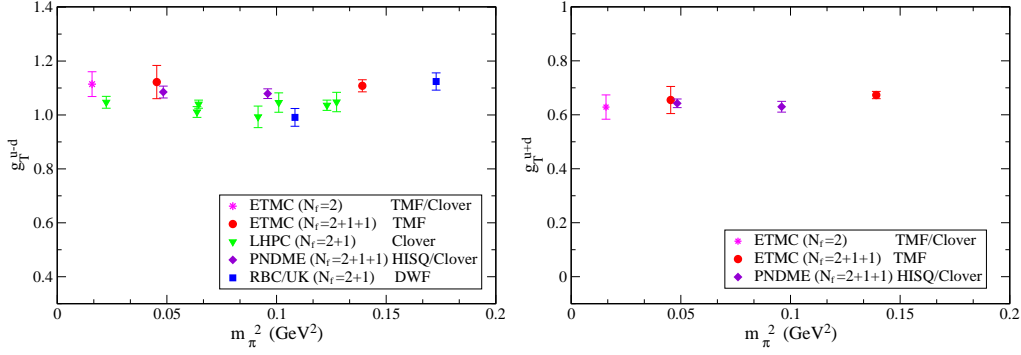

\begin{center}
\includegraphics[scale=0.27]{./gT_vs_mpi_csw.eps}$\,\,$
\includegraphics[scale=0.27]{./gT_vs_mpi_csw_IS.eps}
\end{center}
\caption{The nucleon isovector (left panel) and isoscalar (right panel)
  tensor charge for $N_f{=}2$ TMF with a clover term (magenta
  asterisk) and $N_f{=}2{+}1{+}1$ TMF (red circles), as well as results
  using other lattice actions: green triangles correspond to
  $N_f{=}2{+}1$ clover fermions~\cite{Green:2012ej}, violet diamonds to $N_f{=}2{+}1{+}1$
  clover on HISQ fermions~\cite{Bhattacharya:2013ehc}, blue squares to
  $N_f{=}2{+}1$ domain wall fermions~\cite{Aoki:2010xg} and $N_f{=}2$
  clover fermions~\cite{Pleiter:2011gw}. }
\label{gT_mpi}
\end{figure}

In Fig.~\ref{gT_mpi} we collect our results for the tensor charge.
These are computed at different lattice spacings ranging from $a\sim
0.1$~fm to $a\sim 0.06$~fm, and at different volumes. As can be seen,
there are no sizable cut-off effects. A comparison with other lattice
discretizations~\cite{Aoki:2010xg,Pleiter:2011gw,Green:2012ej,Bhattacharya:2013ehc}
shows that all lattice results are in good agreement.
\begin{figure}[h]
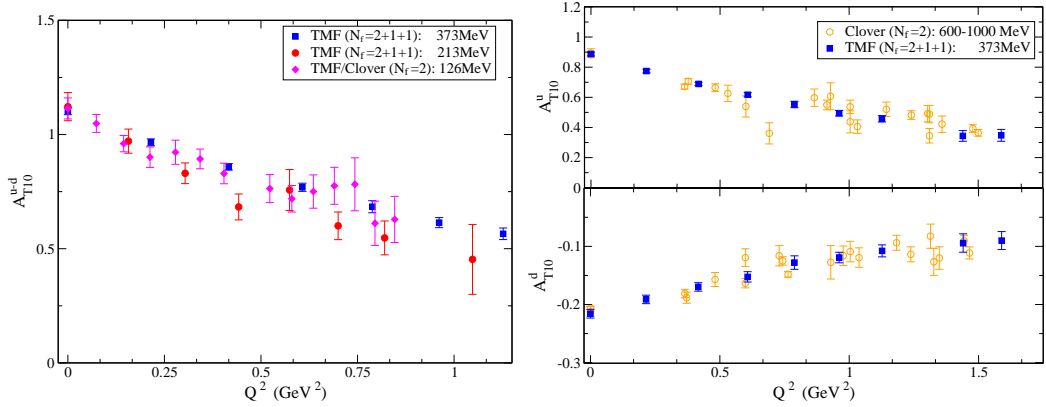

\begin{center}
\includegraphics[scale=0.25]{./AT10_q2_b2.1_b1.95_csw.eps}$\,\,$
\includegraphics[scale=0.25]{./AT10_q2_b1.95_u_d_compare.eps}
\caption{Left panel: The dependence of $A^{u-d}_{T10}$ on the momentum
  transfer, $Q^2$, for i) $N_f{=}2$ TMF with a clover term at $m_\pi=126$~MeV (magenta diamonds), ii) $N_f{=}2{+}1{+}1$ TMF at $m_\pi=213$~MeV
 (red circles) and at $m_\pi=373$~MeV (blue squares). Right panel: a
  comparison between $N_f{=}2{+}1{+}1$ TMF at $m_\pi=373$ MeV
  (blue squares) and $N_f{=}2$ clover fermions at $m_\pi\sim$ 600 -
  1000 MeV (orange circles)~\cite{Pleiter:2011gw}.}
\label{AT10}
\end{center}
\end{figure}
\begin{figure}[h]
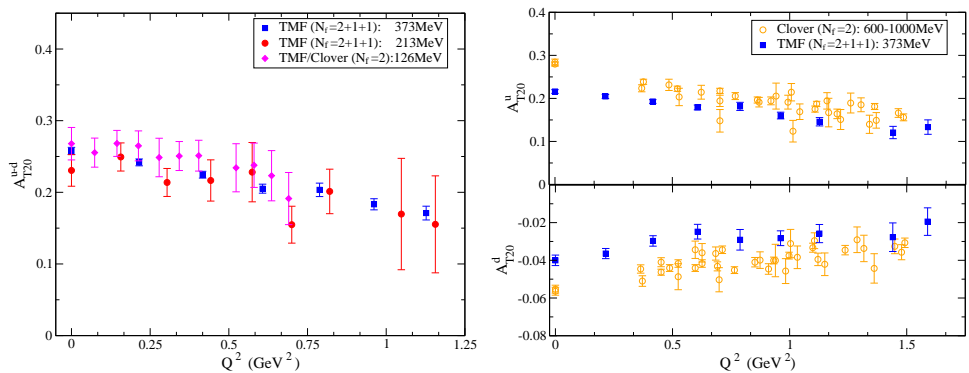

\begin{center}
\includegraphics[scale=0.23]{./AT20_q2_b2.1_b1.95_csw_2.eps}$\,\,$
\includegraphics[scale=0.23]{./AT20_q2_b1.95_u_d_compare.eps}
\caption{$A_{T20}$ versus $Q^2$. The notation is the same as that of Fig. 4.}
\label{AT20}
\end{center}
\end{figure}

In Fig.~\ref{AT10} we show as an example the $Q^2$ dependence of
$A^{u-d}_{T10}$ for the twisted mass results at various pion masses
(left panel) and a comparison with results from $N_f{=}2$ clover
fermions~\cite{Gockeler:2005cj} (right panel). The latter correspond
to a range of values for $m_\pi\sim$ 600-1000 MeV. Despite the
difference in the pion masses, the results
are in good agreement.

From the matrix elements of the one-derivative tensor operator we extract
$A_{T20}$, which is the GFF that can be computed directly from the lattice
data in the forward limit. In Fig.~\ref{AT20} we collect our data for
the isovector case (left panel) and we compare with results from $N_f{=}2$ 
clover fermions~\cite{Gockeler:2005cj} (right panel). Opposed to
$A_{T10}$, we find that $A_{T20}$ is not the same at different
values of the pion mass. This could be due to the perturbative
renormalization and/or a pion mass dependence.

\section{Conclusions}
The tensor charge is evaluated for a range of pion masses including
the physical value. Our values are in agreement with the values
obtained using clover and domain wall fermions. Neglecting
disconnected contributions we find at the physical point
$g^u_T=0.87(4)$ and $g^d_T=0.25(3)$. The first moment of the
transversity distribution is also computed for the first time in the
chiral regime, albeit with a perturbative renormalization. The next
step will be to compute the non-perturbative renormalization for the
transversity distribution.
 
{\bf {Acknowledgments:}} M. C. would like to thank the Cyprus Research
Promotion Foundation for financial support by the project
TECHNOLOGY/$\Theta$E$\Pi$I$\Sigma$/0311(BE)/16. This work used
computational resources provided by PRACE, JSC, Germany.

\end{document}